\documentclass[times]{elsarticle}
\usepackage{framed,multirow}

\usepackage{adjustbox}
\usepackage{amssymb}
\usepackage{latexsym}

\usepackage{stfloats}
\usepackage{dirtytalk}

\usepackage{url}
\usepackage{xcolor, soul, color}
\usepackage{lipsum}  
\usepackage{amsmath}
\usepackage{tabularx}

\usepackage{hyperref,tabularx,graphicx, multirow,subcaption}
\usepackage{newtxmath, mleftright}
\usepackage{cleveref}
\definecolor{newcolor}{rgb}{.8,.349,.1}

\journal{Medical Image Analysis}

\begin{document}

%\verso{Alessandro Casella \textit{et~al.}}

\begin{frontmatter}

\title{Learning-Based Keypoint Registration for Fetoscopic Mosaicking}
%\title{Type the title of your paper, only capitalize first word and proper nouns\tnoteref{tnote1}}%
%\tnotetext[tnote1]{This is an example for title footnote coding.}

\author[2,3]{Alessandro Casella\corref{cor1}}
\cortext[cor1]{Corresponding author.}
\ead{alessandro.casella@iit.it}% alessandro.casella@polimi.it
\author[1]{Sophia Bano}
\author[1]{Francisco Vasconcelos} %f.vasconcelos@ucl.ac.uk
\author[14,15,16]{Anna L. David}
\author[17]{Dario Paladini}
\author[15,16]{Jan Deprest}
\author[3]{Elena De Momi} % elena.demomi@polimi.it
\author[2]{Leonardo S. Mattos} %
\author[4]{Sara Moccia} %
\author[1]{Danail Stoyanov}

%\author[2]{Given-name4 \snm{Surname4}}

\address[1]{Wellcome/EPSRC Centre for Interventional and Surgical Sciences (WEISS) and Department of Computer Science, University College London, UK}
\address[2]{Department of Advanced Robotics, Istituto Italiano di Tecnologia, Italy}
\address[3]{Department of Electronics, Information and Bioengineering, Politecnico di Milano, Italy}
\address[14]{Fetal Medicine Unit, Elizabeth Garrett Anderson Wing, University College London Hospital, UK}
\address[15]{EGA Institute for Women's Health, Faculty of Population Health Sciences, University College London, UK}
\address[16]{Department of Development and Regeneration, University Hospital Leuven, Belgium}
\address[17]{Department of Fetal and Perinatal Medicine, Istituto "Giannina Gaslini", Italy}
\address[4]{The BioRobotics Institute and Department of Excellence in Robotics and AI, Scuola Superiore Sant’Anna, Italy}

%\received{Day Month 2022}
%\finalform{Day Month 2022}
%\accepted{Day Month  2022}
%\availableonline{Day Month  2022}
%\communicated{S. Sarkar}

\begin{abstract}
In Twin-to-Twin Transfusion Syndrome (TTTS), abnormal vascular anastomoses in the monochorionic placenta can produce uneven blood flow between the two fetuses. In the current practice, TTTS is treated surgically by closing  abnormal anastomoses using laser ablation. This surgery is minimally invasive and relies on fetoscopy. Limited field of view makes anastomosis identification a challenging task for the surgeon.
% 
%\textbf{Methods:} 
To tackle this challenge, we propose a learning-based framework for in-vivo fetoscopy frame registration for field-of-view expansion. The novelties of this framework relies on a learning-based keypoint proposal network and an encoding strategy to filter (i) irrelevant keypoints based on fetoscopic image segmentation and (ii) inconsistent homographies.
% 
%\textbf{Results:} 
We validate of our framework on a dataset of 6 intraoperative sequences from 6 TTTS surgeries from 6 different women against the most recent state of the art algorithm, which relies on the segmentation of placenta vessels.
% 
%\textbf{Conclusion:} 
The proposed framework achieves higher performance compared to the state of the art, paving the way for robust mosaicking to provide surgeons with context awareness during TTTS surgery.
\end{abstract}


\end{frontmatter}

%\linenumbers

%\hl{Structure: 

%(1) Provide a detail review of existing SOA in fetoscopic segmentation and registration, 

%(2) Challenge overview including introduicng the data, challenge tasks, submission protocol, evaluation metrics etc, (3) 

%Team Submissions (to include a brief summary of all teams and preferably have a method overview figure for each team), 

%(4) Results and discussion for Task 1 and Task 2.}

\section{Introduction}\label{intro}

Twin-to-twin Transfusion Syndrome~(TTTS) is a rare complication affecting 10-15\% of monochorionic diamniotic pregnancies. TTTS is characterized by the development of unbalanced and chronic blood transfer from one twin (the donor) to the other (the recipient), through placental communicating vessels called anastomoses~\citep{Baschat2011}.
This shared circulation causes profound fetal hemodynamic unbalance and consequently severe growth restriction, cardiovascular dysfunction, hypoxic brain damage and death of one or both twins~\citep{Lewi2013}.

\begin{figure*}[t!]
    \centering
   \includegraphics[width=\textwidth]{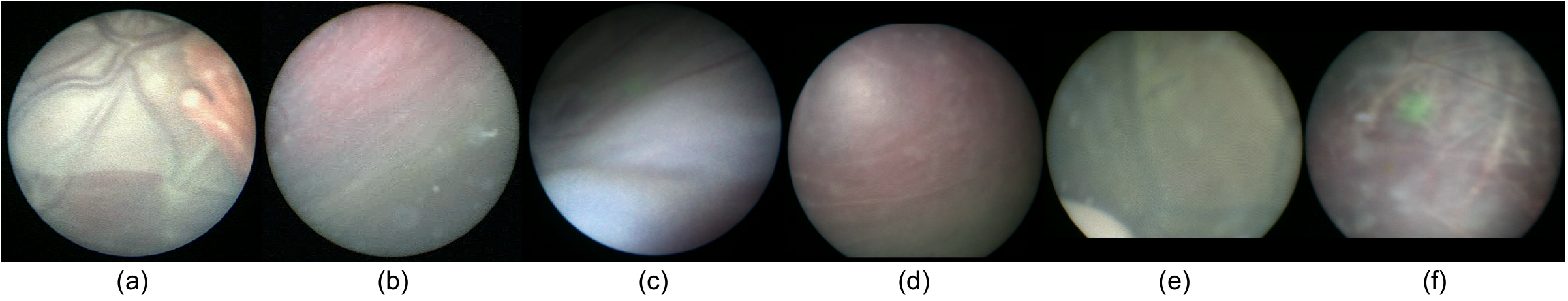}
    \caption{Main challenges of TTTS frames: (a) occlusions, (b, d, f) lack of anatomical structures (e.g., vessels), (c) poor frame texture,  (e)  non-planar view in case of anterior placenta.}
    \label{fig:challenge}
\end{figure*}

The recognized elective treatment for TTTS is selective laser photocoagulation of anastomoses originating in the donor’s placental territory. This treatment requires precise identification and laser ablation of placental vascular anastomoses~\citep{beck2012preterm}. %It relies on the classification of anastomoses in arterio-venous (from donor to recipient, AVDR, or from recipient to donor, AVRD), arterio-arterial (AA) or veno-venous (VV), and may distribute themselves following a certain pattern on the placenta. 
%
%Recent researches have identified that the sequence in which these anastomoses are treated with the laser could result in further hypotension of the donor twin, with increased risks of complications and fetal demise~\cite{Nakata2009}. 
%
%Up to now, significant complication or failures were recorded in both techniques, the reason being that tiny anastomoses might be overlooked because flattened on the placental surface by the high pressure in the recipient’s sac. Therefore, the Solomon technique was introduced~\cite{Slaghekke2016} and now has become the gold standard in those cases in which it can be technically achieved. This technique consists in drawing a coagulation line connecting all the sites in which anastomoses had been coagulated after the first selective ablation round.
%
Despite recent advancements in instrumentation and imaging for TTTS~\citep{Cincotta2016}, residual anastomoses after ablation still represent a major complication~\citep{Lopriore2007}.  % in monochorionic placentas treated with fetoscopic laser surgery~\cite{Lopriore2007}. 
This may be explained considering the challenges, from the surgeon's side, of limited Field of View (FoV) and constrained manoeuvrability of the fetsocope, especially for anterior placenta. % due to the unfavourable viewing angle.% Main image challenges as shown in Fig.~\ref{fig:challenge}. 
%potrebbe esserci placenta anteriore, ma non sono ipersicuro 
%\textcolor{red}{(Fig.~con challenge)} 
%
In this complex scenario, Computer-Assisted Intervention (CAI) and Surgical Data Science (SDS) methodologies~\citep{maier2022surgical} may be exploited to provide surgeons with mosaicking for FoV expansion. 

An approach to mosaicking relying on external devices is proposed in~\cite{tella2016combined}.
However, external devices may not be always used in the operating room due to current regulations. 
Currently, researchers are focusing on methods to perform mosaicking using only fetoscopy images. Several challenges in endoscopic images hamper the translation of previously developed methods into the actual surgical practice, as reported by~\cite{9268990}. These challenges include poor visibility due to amniotic fluid turbidity, low-resolution of fetoscopic images, occlusions by surgical tools and fetus (Fig.~\ref{fig:challenge} (a)), lack of anatomical structures (Fig.~\ref{fig:challenge} (b, d, f)) to be used as reference for frame registration, poor frame texture (Fig.~\ref{fig:challenge} (c)) and distortion introduced by non-planar views due to fetoscope camera orientation, especially in case of anterior placenta (Fig.~\ref{fig:challenge} (c, e)).~\cite{bano2021fetreg}
%The potential benefit of addressing reduced visibility in fetoscopy has focused researchers' attention on mosaicking techniques for expanding the field of view.
%
%

First approaches to mosaicking using only fetoscopy images, include the work of~\cite{daga2016real} and \cite{reeff2006mosaicing}, who relied on standard descriptors, such as SIFT~(\cite{Gutirrez2016SIFTI}) and SURF~(\cite{10.1007/11744023_32}). These methods were validated on synthetic phantoms or ex-vivo placental sequences only, and may fail when processing in-vivo placenta frames, due to the lack of texture typical of intra-operative endoscopy images.
The approach in~\cite{peter2018retrieval} follows a different approach, minimizing the photometric consistency between frames, showing promising results with in-vivo fetoscopy data. However, the computation time to process a frame pair is a major bottleneck and may not be compatible with real-time mosaicking.

More recently, deep-learning algorithms have been proposed to tackle the challenges of fetoscopy frames. In~\cite{gaisser2018stable}, stable regions manually identified in the frames are used as a prior for frame registration with a convolutional neural network. The approach is tested on phantoms only.
The work in~\cite{bano2019deep} uses HomographyNet~\citep{detone2017toward} to perform pair-wise homography estimation, but the validation is performed on a single in-vivo sequence. 

A recent and promising approach in the field, which is presented in~\cite{bano2020deep,bano2020deepijcars}, shows that placental vessels provide unique landmarks to compute homography. %, overcoming the limits in~\cite{bano2019deep} by being robust to fetus presence in the FoV.
While obtaining accurate vessel segmentation might be considered an affordable challenge~\citep{bano2021fetreg}, the approach in~\cite{bano2020deep,bano2020deepijcars} fails whenever vessels are not visible. % as in Fig.~\ref{fig:challenge}. 

%More recently,the work in~\cite{tosin2022robust} proposes a method based on FlowNet-v2~\citep{ilg2017flownet} and RANSAC~\citep{Fischler1981} for computing the motion field and relative homography between adjacent fetoscopy frames. 
%Unlike~\cite{bano2020deep,bano2020deepijcars}, this method does not rely on vessel segmentation, achieving mosaicking even if vessels are not present or clearly visible. However, FlowNet-v2 assumes brightness constancy and strong texture~\citep{Shah2021}, which cannot be always guaranteed in fetoscopic frames. 

%

\begin{figure*}[t!]
    \centering
    %\makebox[\columnwidth]{
    \includegraphics[width=.7\textwidth]{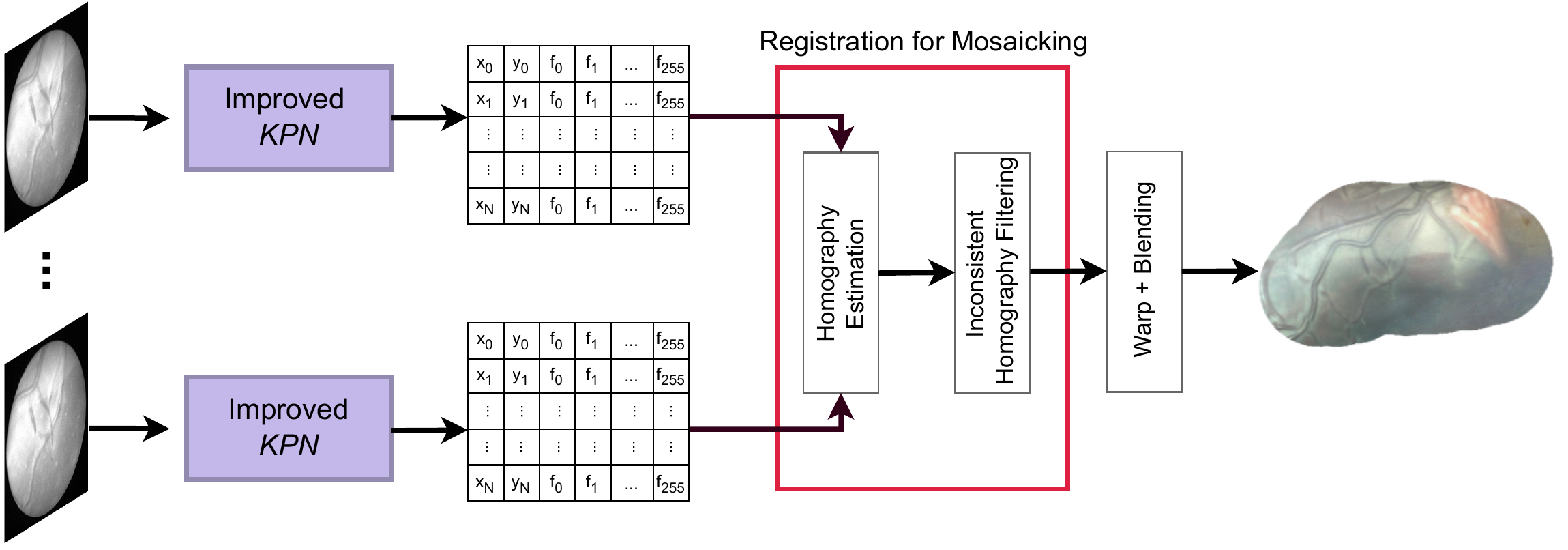}
    \caption{Overview of our mosaicking framework. The Keypoint Proposal Network (\textit{KPN}) computes keypoints that are then filtered to reject irrelevant keypoints. Registration for mosaicking is performed to register consecutive fetoscopy frames. Warping and blending are performed for visual purposes. }
    \label{fig:overall}
\end{figure*}

The challenges of in-vivo fetoscopy video analysis may explain why only few researchers have attempted to design algorithms for fetoscopy mosaicking so far. Last year, we organized the sub-challenge \say{FetReg: Placental Vessel Segmentation and Registration in Fetoscopy}\footnote{\url{https://fetreg2021.grand-challenge.org/}}, inside EndoVis, a MICCAI Grand Challenge. 
Only one team competed to the task \say{Placental vessel registration and RGB frame registration for mosaicking.}~\cite{fetreg2021v2}

With this work, we aim to contribute to the advancement of the state of the art in fetoscopy mosaicking by investigating,  with a comprehensive study with 6 videos (14500 frames) acquired from 6 women during actual surgery, the following research hypotheses:

\begin{itemize}
    \item Hypothesis 1 (H1): Keypoint learning can tackle the challenges typical of fetoscopic videos acquired during TTTS surgery and provide robust keypoints for mosaicking without relying on the segmentation of structures in the FoV % neither assuming constant illumination.

    \item Hypothesis 2 (H2): Mosaicking performance can be boosted by filtering irrelevant keypoints and rejecting inconsistent homographies.
\end{itemize}

\subsection{Contribution}
In this paper, we propose a learning-based framework for the robust detection of keypoints with the aim to register consecutive fetoscopy images acquired during TTTS surgery and accomplish fetoscopy mosaicking.
Our framework does not rely on any structure segmentation for keypoint estimation. However, when fetus and surgical tools are in the FoV, their segmentation is used for irrelevant keypoint rejection. 

%
%To experimentally validate our framework, we rely on the dataset published in~\cite{bano2019deep}, which is, so far, the only publicly available one for mosaicking. %The dataset includes 6 videos acquired during the actual surgical practice from 6 cases of TTTS.
The contributions of this work can be summarized as follows:
\begin{enumerate}
    \item Development of a new framework for robust FoV expansion in TTTS fetoscopy videos, which features an intrinsic strategy for detecting  keypoints robustly and filtering inconsistent homographies.
    \item Design of a self-supervised strategy for training the framework with unlabeled fetoscopy frames.
    %that outperforms the state of the art when tested %on in-vivo fetoscopic images~\cite{bano2020deep}.
    %\item A keypoint rejection strategy relying on automatic structure segmentation for filtering irrelevant keypoints % could affect homography estimation.
   % \item An homoghraphy filter that discard frames with unrealistic camera motion estimation.
    %\item Introduction of an irrelevant keypoint rejection strategy that relies on automatic segmentation of fetus and tool.
    %\item Validation of fetoscopic image registration using keypoints computed on features learned on fetoscopic data. We show that this method is reliable even when vessels are not clearly visible.
    \item Validation using the largest dataset in the field, which consists of 6 in-vivo TTTS videos from~\cite{bano2019deep}.
    %\item A quantitative comparison of the registration consistency in respect to available alternatives by analysing the similarity between overlapping frames. 
\end{enumerate}

\begin{figure*}[t!]
    \centering
    \includegraphics[width=.8\textwidth]{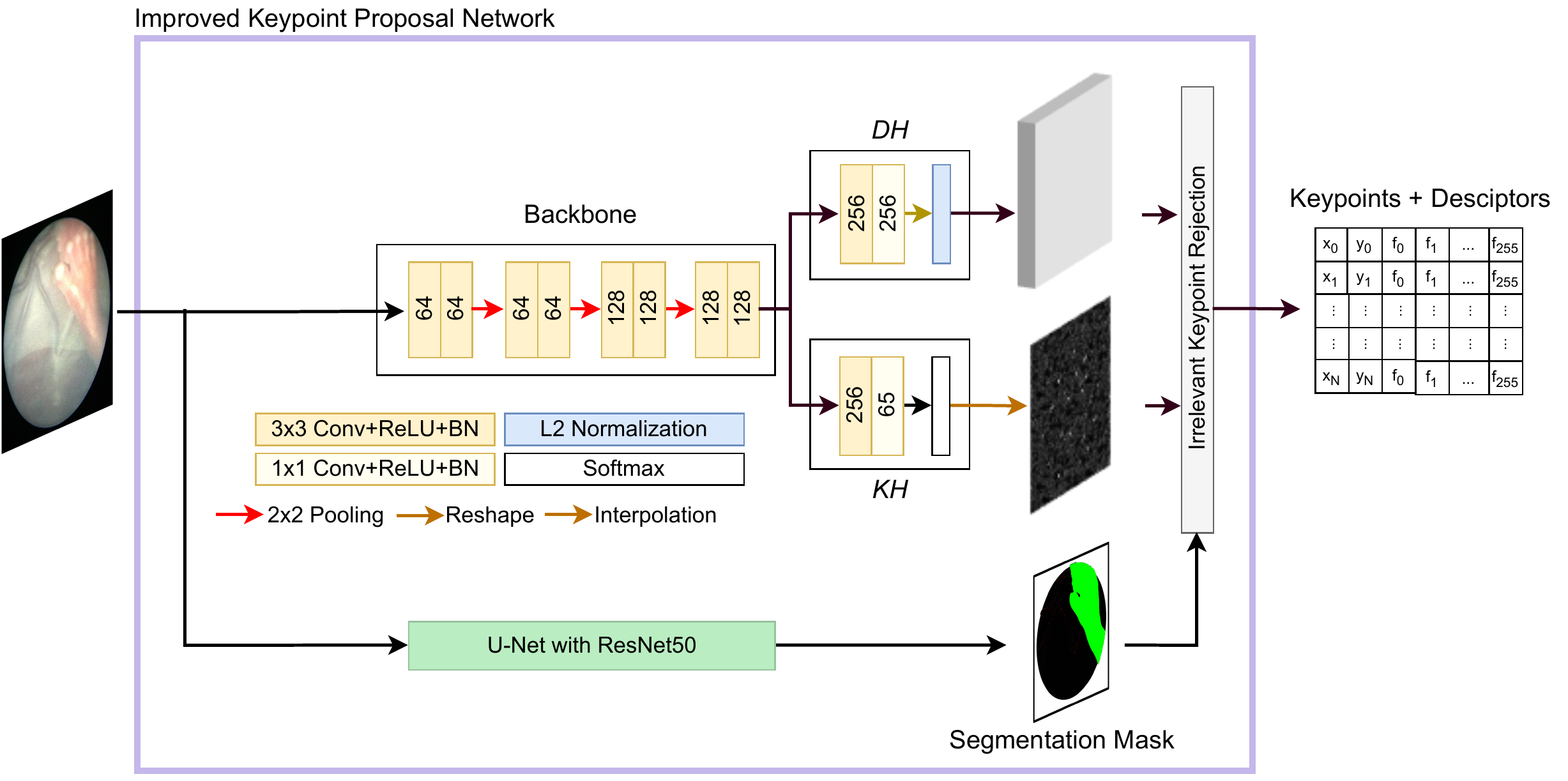}
    \caption{Overview of the improved Keypoint Proposal Network.
    \textit{KH} and \textit{DH} are the keypoint and keypoint descriptor head, respectively.
    Irrelevant keypoint rejection relies on surgical-tool and fetus segmentation performed by the U-Net with ResNet50 backbone from~\cite{bano2020deepijcars}. The overall output is a set of keypoints and their descriptors. % registration framework. The Keypoint Proposal Network computes keypoints that are then filtered by the \textit{Keypoint Filter} to discard invalid and not useful keypoints. Filtered keypoints are then processed by the \textit{Registration and Blending} block that compute the transformation between the keypoints of consecutive frames and perform the blending. \textcolor{red}{we want to see  keypoint proposal computation, irrelevant keypoint rejection, and homography estimation and filtering }
    }
    \label{fig:KPN}
\end{figure*}

To the best of our knowledge, this work is the first to investigate the  potential of keypoint learning for fetoscopy mosaicking. We perform extensive comparison with the state of the art, as well as an ablation study to identify the best configuration of our framework.

\section{Method}\label{methods}

Our framework consists of improved Keypoint Proposal Network (\textit{KPN}) (Sec.~\ref{sec:improved_KPN}) for keypoints learning keypoints and irrelevant keypoint rejection, 
and registration for mosaicking (Sec.~\ref{sec:registr}), which estimates homography from the keypoints and filters inconsistent homographies. {The overall framework is shown in Fig.~\ref{fig:overall}. }

\subsection{Improved Keypoint Proposal Network} \label{sec:improved_KPN}

The architecture of the improved \textit{KPN} is shown in Fig.~\ref{fig:KPN} and consists of keypoint proposal computation (Sec.~\ref{kpc}), using the \textit{KPN} trained following the strategy described in Sec.~\ref{kpn_train}, and irrelevant keypoint rejection (Sec.~\ref{kfilt}). 

\subsubsection{Keypoint proposal computation.}  \label{kpc}

The \textit{KPN}, is a convolutional neural network based on SuperPoint proposed by~\cite{detone2018superpoint}.
SuperPoint is the learning-based keypoint detection network that exibits state-of-the-art performance on a large number of geometry problems in computer vision, including homography estimation, where groundtruth is not available.

\textit{KPN} consists of a VGG-16 backbone for feature extraction, followed by two heads, the Keypoint Head (\textit{KH}) for the detection of candidate keypoints, and the Descriptor Head (\textit{DH}) for computing keypoint descriptors.
\textit{KH} outputs a dense point map, with the same size as the input frame, where the value of each pixel refers to the probability of that pixel of being a keypoint.
%Teach pixel corresponds to a candidate keypoint, and its value is associated with the confidence of that candidate to be a keypoint.
%
\textit{DH} outputs a {L2}-normalized descriptor vector for each candidate keypoint. %This vector has size equal to 256 to be consistent with the first convolution block of the head. +++

\subsubsection{{KPN training.}} \label{kpn_train}
We train the \textit{KPN} in four steps, taking inspiration from~\cite{detone2018superpoint}.
As a first step, to account for the lack of annotated TTTS frames, we train \textit{KPN} without \textit{DH} on a synthetic dataset. 
We use a synthetic image-keypoint pairs obtained from the synthetic shapes dataset presented in~\cite{detone2017toward}. Each pair consists of images with size $448x448$ pixels containing simple polygons and associated keypoints. 
To increase the variability of the dataset, we apply during training (i) perspective distortions (i.e. homographic augmentation), to model different camera views, and (ii) brightness and contrast augmentation (i.e. photometric augmentation), to encode the intensity variability.
%During training, we perform homography-based augmentation, similarly to what is done in . %Each pair and transformation is generated on-the-fly during the training, providing high variability of keypoints. %

As a second step, we fine tune \textit{KPN}, still without \textit{DH}, on natural images from MS-COCO 2014 training dataset~(\cite{10.1007/978-3-319-10602-1_48}). In this case, we follow a self-supervised training strategy to account for the lack of keypoint annotation in the dataset.
We infer the MS-COCO 2014 test dataset using the weights obtained in the first step. From each image in the dataset, a patch of $448x448$ pixels is randomly cropped and converted to grayscale. The estimated dense point map is used to generate the pseudo-ground truth.
Also in this case, photometric and homography-based augmentation is applied on the fly. 
%The first one includes classical perturbations such as Gaussian noise, motion blur, random brightness and contrast, while the latter is added to enforce the proposal of keypoints consistent with a defined set of plausible camera transformations. Both transformations are applied during the training.
%%% supplementary %%%%

%%% supplementary %%%%

The last two steps involve the TTTS dataset.
In the third step, we infer a subset of our TTTS dataset using the weights learned at the second step to obtain the associated pseudo-ground truth. We use this pseudo-ground truth to fine-tune the \textit{KPN} without \textit{DH}.
In the last step, we update the TTTS-image pseudo-ground truth using the weights obtained at the third step. This pseudo-ground truth is used to train the whole \textit{KPN} on TTTS images. For homographic augmentation, we limit the parameter range to be consistent to fetoscope camera model.
%

% inlcude the loss from superpoint

For all steps, we use the loss function $\mathcal{L}_{\mathit{KPI}}$ defined as: 
\begin{equation}
    \mathcal{L}_{\mathit{KPI}}=\mathcal{L}_{KP}+\mathcal{L}_{KP}'+\lambda\mathcal{L}_{D}(D,D')
\end{equation}
$\mathcal{L}_{KP}$ is the cross-entropy loss computed over the keypoint map generated by \textit{KH} and its groundtruth, $\mathcal{L}_{KP}'$ is the loss computed on the warped keypoint map generated by \textit{KH} after image warping with a random homography, while $\mathcal{L}_{D}(D,D')$ is the hinge loss between descriptors from the original image and those from the warped image weighted by the term $\lambda$. $\lambda$ is adjusted during training to balance the effect $\mathcal{L}_{D}(D,D')$ term that, especially in the first training epochs, has largely negative values.

\subsubsection{{Irrelevant keypoint rejection.}} \label{kfilt}

We noticed experimentally that the \textit{KPN} finds keypoints also on structures, such as fetus and surgical tools, that are not relevant to model fetoscope movement. These keypoints could affect homography estimation negatively. To reject irrelevant keypoints, we filter out keypoint proposals that fall inside fetus and surgical tool segmentation masks. These masks are obtained using the U-Net with ResNet50 backbone  model presented in~\cite{bano2021fetreg}.

\subsection{Registration for Mosaicking} \label{sec:registr}

\subsubsection{{Homography estimation.}}
%that could led to wrong registration, we provided a binary mask \textcolor{red}{of fetus and surgical tool} to perform a first filtering of unwanted keypoints. % (e.g. outside the Field of View, Fetus or Tool).
Assuming \textit{KPN} to be robust, we design a simple frame-pair registration pipeline to achieve fast registration at low computational cost. 
We approximate registration as affine transformations, assuming that this can provide a reasonably good description of fetoscope camera movement, following considerations in~\cite{bano2020deep}.
The homography of two consecutive frames is estimated using Levenberg-Marquardt optimisation. 
RANSAC is used to find keypoints that match affine transformation constraint. 
Other than RANSAC, we also evaluated the use of other homography estimation algorithms, such as PROSAC and MAGSAC. However, a comparison of these methods in a preliminary analysis did not show any signifcant difference in the mosaicking performance. 
%
%According to the observations in~\cite{bano2020deep}, projective registration leads to poor registration performance. The higher degrees of freedom, the lack of camera calibration and lens distortion that cannot be compensated will lead to the computation of inconsistent homographies. 

\begin{table*}[t!]
\centering
\caption{Characteristics of the dataset presented in~\cite{bano2020deep}.}
\begin{adjustbox}{max width = \textwidth}
\begin{tabular}{cccc}
Video ID & Frame number & Frame resolution & Placenta position  \\
&& [pixels]&\\
        \hline
 1 & 400       & 470x470                   & Posterior \\
 2 & 300       & 540x540                   & Posterior \\
 3 & 150       & 550x550                   & Anterior  \\
 4 & 200       & 640x640                   & Posterior \\
 5 & 200       & 720x720                   & Anterior  \\
 6 & 200       & 720x720                   & Posterior
\end{tabular}
\end{adjustbox}
\label{tab:dataset}
\end{table*}

\subsubsection{{Inconsistent homography filtering.}}
We can assume that homography should not reflect large displacement, rotation or scaling, as we register consecutive frames. %Any such case is considered as an incorrect estimate or inconsistent homography.  
%To handle this inconsistency, 
We take inspiration from~\cite{bano2019deep} to filter out any homography that does not reflect this assumption.
We perform singular value decomposition on each estimated homography to extract rotation, scale and translation parameters. 
When one of these parameters exceeds a threshold defined experimentally, the second frame in the pair to be registered is discarded, and the registration with the next frame is attempted. 
%Thresholds are chosen experimentally according to previous observations.
This procedure is reiterated for five frames and, in case of failure, mosaicking computation ends.

%Registration is aborted if registration fails after the fifth successive frame.
%We have noticed that some strange transformations can occur in some frame pairs, introducing drift or causing mosaicking to fail. This is reflected in a large change of values in the homography, in particular in the upper left submatrix that describes the scale and the rotation. 
%To handle this issue, we perform singular value decomposion to extract the components of the  motion. When a frame is discarded, we attempt registration with the next frame in order not to interrupt mosaicking. Registration is aborted if registration fails after the fifth successive frame.

\section{Experimental Setup}\label{table:dataset}

\subsection{Dataset}\label{sec:dataset}

We validate our framework using an extended version of the dataset published in~\cite{bano2020deep}.
The overall dataset consists of 1450 frame from 6 different in-vivo TTTS fetoscopy procedures. Main characteristics of the dataset are summarized in Table~\ref{table:dataset}. Frame number and resolution varies from video to video. %Videos of anterior and posterior placenta are represented in the dataset.

Videos differ in terms of intra-operative environment, artifacts and lighting conditions. Two videos present anterior placenta. While in posterior-placenta videos, the scene can be considered nearly planar because a straight fetoscope is used, with anterior placenta, the surgeon rely on the 30-degrees fetoscope~(\cite{awais,casella2021shape}). The resulting non-planar view adds further challenges to mosaicking, as introduced in Sec.~\ref{intro}.

\begin{table*}[t!]
    \centering
        \caption{Ablation study - summary. \textit{KPN}: Keypoint proposal network (Sec.~\ref{kpc}).  $^*$ For E0, we test SuperPoint trained on MS-COCO 2014 dataset without any fine tuning on fetoscopy data.}
    \label{tab:abl}
    \begin{adjustbox}{max width = \textwidth}
    \begin{tabular}{cccc}
        & \textit{KPN} & Irrelevant keypoint & Inconsistent homography  \\
        &  & rejection & filtering\\
        \hline
        E0 & X$^*$ & & \\
        E1 & X & & \\
        E2 & X  & X & \\
        Proposed & X  & X & X\\
    \end{tabular}
    \end{adjustbox}
\end{table*}

\begin{table*}[t!]
\centering
\caption{Quantitative results for the 6 tested in-vivo fetoscopy videos. The $s$ with \textit{n} = 5 frames is reported in terms of mean $\pm$ standard deviation.}  \label{tab:res}
\begin{tabular*}{.6\textwidth}{l@{\extracolsep{\fill}}ccc}
                           & Video 1      & Video 2      & Video 3      \\ \hline
\multicolumn{1}{l}{SIFT} & \multicolumn{1}{c}{$0.662 \pm 0.115$} & \multicolumn{1}{c}{$0.732 \pm 0.120$} & \multicolumn{1}{c}{$0.749 \pm 0.279$} \\ 
\multicolumn{1}{l}{\cite{bano2020deep}} &
  \multicolumn{1}{c}{$\mathbf{0.757 \pm 0.081}$} &
  \multicolumn{1}{c}{$\mathbf{0.788 \pm 0.050}$} &
  \multicolumn{1}{c}{$0.839 \pm 0.208$} \\
  \multicolumn{1}{l}{Pre-trained SuperPoint (E0)} & \multicolumn{1}{c}{$0.528 \pm 0.247$} & \multicolumn{1}{c}{$0.202 \pm 0.264$} & \multicolumn{1}{c}{$0.219 \pm 0.266$} \\ 
\multicolumn{1}{l}{Vanilla SuperPoint (E1)} & 
 \multicolumn{1}{c}{$0.731 \pm 0.116$} &
  \multicolumn{1}{c}{$0.740\pm 0.079$} &
  \multicolumn{1}{c}{$0.809 \pm 0.174$}  \\
\multicolumn{1}{l}{Improved $KPN$ (E2)} & 
 \multicolumn{1}{c}{$0.730 \pm 0.112$} &
  \multicolumn{1}{c}{$0.743\pm 0.071$} &
  \multicolumn{1}{c}{$0.813 \pm 0.172$} \\ 
\multicolumn{1}{l}{Proposed} & 
  \multicolumn{1}{c}{$0.750 \pm 0.081$} &
  \multicolumn{1}{c}{$0.766 \pm 0.048$} &
  \multicolumn{1}{c}{$\mathbf{0.884 \pm 0.075}$} \\ \\
                           & Video 4      & Video 5      & Video 6      \\ \hline
\multicolumn{1}{l}{SIFT} & \multicolumn{1}{l}{$0.660 \pm 0.347$} & \multicolumn{1}{l}{$0.5164 \pm 0.402$} & \multicolumn{1}{l}{$0.485 \pm 0.389$} \\
\multicolumn{1}{l}{\cite{bano2020deep}} &
  \multicolumn{1}{c}{$0.745 \pm 0.257$} &
  \multicolumn{1}{c}{$0.890 \pm 0.070$} &
  \multicolumn{1}{c}{$0.861 \pm 0.205$} \\ 
\multicolumn{1}{l}{Pre-trained SuperPoint (E0)} & \multicolumn{1}{c}{$0.322 \pm 0.362$} & \multicolumn{1}{c}{$0.341 \pm 0.284$} & \multicolumn{1}{c}{$0.209 \pm 0.336$} \\ 
\multicolumn{1}{l}{Vanilla SuperPoint (E1)} & 
 \multicolumn{1}{c}{$0.801\pm 0.111$} &
  \multicolumn{1}{c}{$0.829\pm 0.091$} &
  \multicolumn{1}{c}{$0.817 \pm 0.076$}  \\
\multicolumn{1}{l}{Improved $KPN$ (E2)} &
 \multicolumn{1}{c}{$0.818 \pm 0.111$} &
  \multicolumn{1}{c}{$0.832 \pm 0.090$} &
  \multicolumn{1}{c}{$0.817 \pm 0.073$} \\
\multicolumn{1}{l}{Proposed} &
  \multicolumn{1}{c}{$\mathbf{0.870 \pm 0.125}$} &
  \multicolumn{1}{c}{$\mathbf{0.897 \pm 0.012}$} &
  \multicolumn{1}{c}{$\mathbf{0.909 \pm 0.021}$} \\
\end{tabular*}
\end{table*}

\subsection{Implementation details}

Our framework is implemented in TensorFlow 1.15 and trained on two NVIDIA A100 40GB, using ADAM optimizer and a learning rate of $10^{-3}$. 
For training the \textit{KPN} following the strategy described in Sec.~\ref{kpn_train}, in the first 3 training steps we set a batch size of 64, while a batch size of 8 is used for the last step. For the 4 steps, we set a number of iteration equal to 180000, 60000, 20000 and 12000, respectively.

\subsection{Performance metrics}

We measure the performance of our framework using the structural similarity index measure (SSIM) over a number ($n$) of frames, with $n \in [1, 5]$, for fair comparison with~\cite{bano2020deep}. We call this metric $s$.
Given a source ($I_{i}$) and a target ($I_{i+n}$) frame, and a homography transformation (${H}_{i \rightarrow i+n}$) between $I_{i}$ and $I_{i+n}$, for every $i$-th frame in the TTTS sequence $s$ is defined as:
\begin{equation}
    s_{i \rightarrow i+n} =\mathrm{sim}(w(\tilde{I}_{i},H_{i \rightarrow i+n}),\tilde{I}_{i+n})
\end{equation}
where $\mathrm{sim}()$ is the standard formula for SSIM, $w$ is the warping operator, and $\tilde{I}$ and $\tilde{I}_{i+n}$ are smoothed versions of $I$ and $I_{i+n}$, respectively.
$\tilde{I}$ and $\tilde{I}_{i+n}$ are obtained by applying $9\times9$ Gaussian filtering with standard deviation of $1.5$. This makes $s$ robust even in presence of amniotic fluid particles and fetoscopy-image noise. %
%The $s$ is computed considering only the central area of TTTS frames, excluding the surrounding black border. 
When exploring the vascular network, the fetoscope mainly makes small movements. Using similarity metrics with images with low texture whose overlap percentage is high due to small displacements results in very high similarity values that are not useful for identifying differences between fetoscopic image registration methods. %
Using  values of $n$ larger than 1 allows us to assess the presence of drift in consecutive homographies. % of the chain of homographs, and therefore the goodness of the registration algorithm, to be taken into account and a higher probability that the frames overlap imperfectly. In the case of sudden movements, this condition may not be maintained.
%For this reason, we chose to set $n=5$ to have a sufficient distance to have different frames whose overlaps, but the probability that the frames would be non-overlapping due to very large movements would be almost zero

\begin{figure*}[t!]
\centering
    \makebox[\textwidth]{
\begin{tabular}{cc}
  \includegraphics[width=0.45\textwidth]{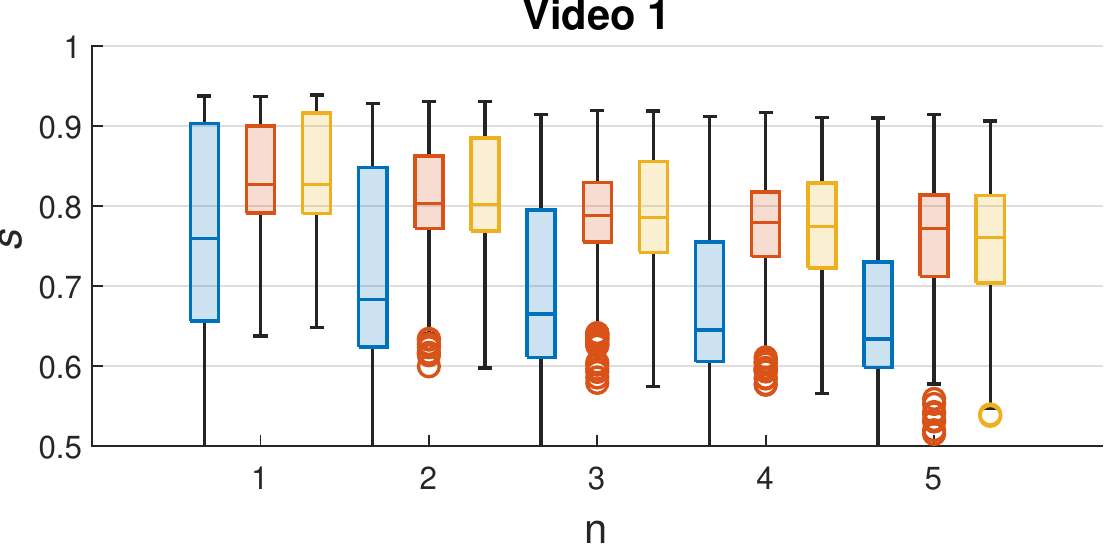} &   \includegraphics[width=0.45\textwidth]{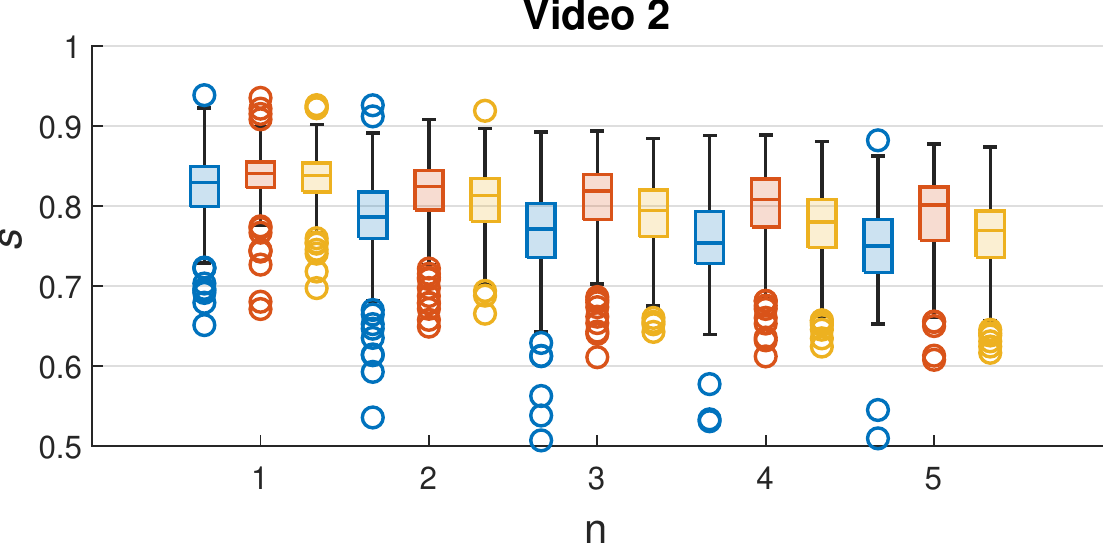} \\ \includegraphics[width=0.45\textwidth]{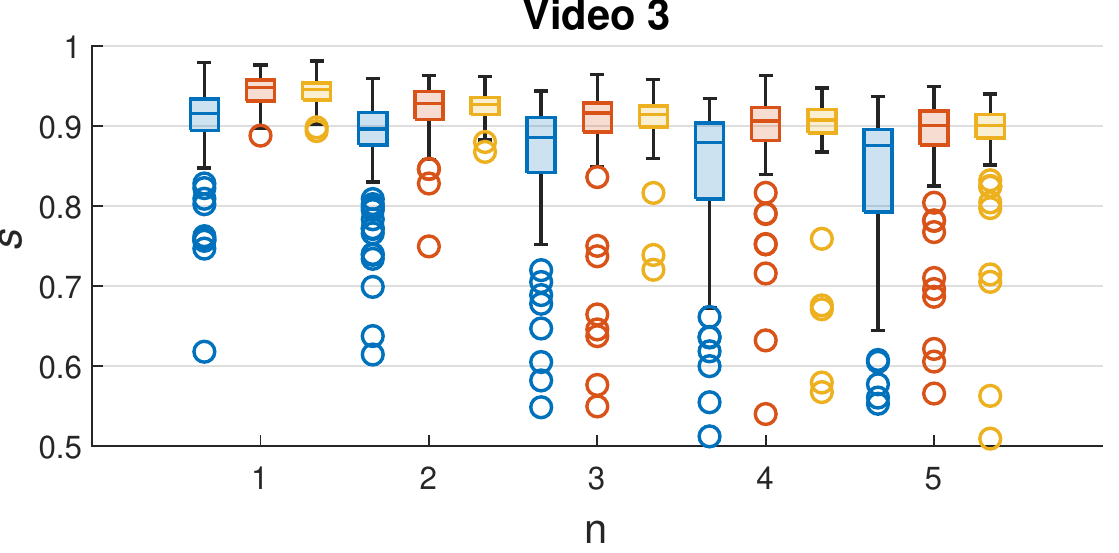} & \includegraphics[width=0.45\textwidth]{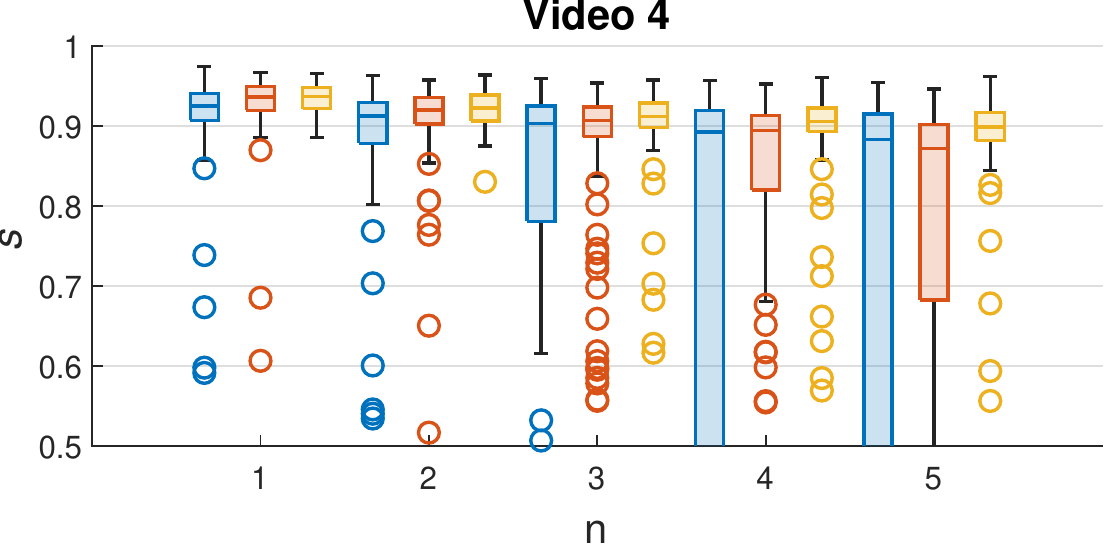} \\   \includegraphics[width=0.45\textwidth]{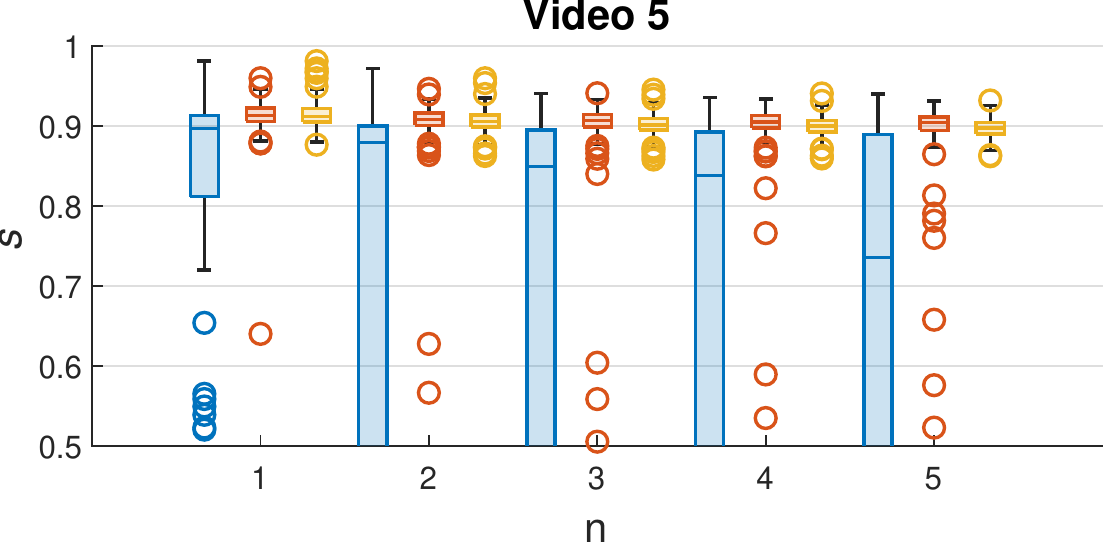} & \includegraphics[width=0.45\textwidth]{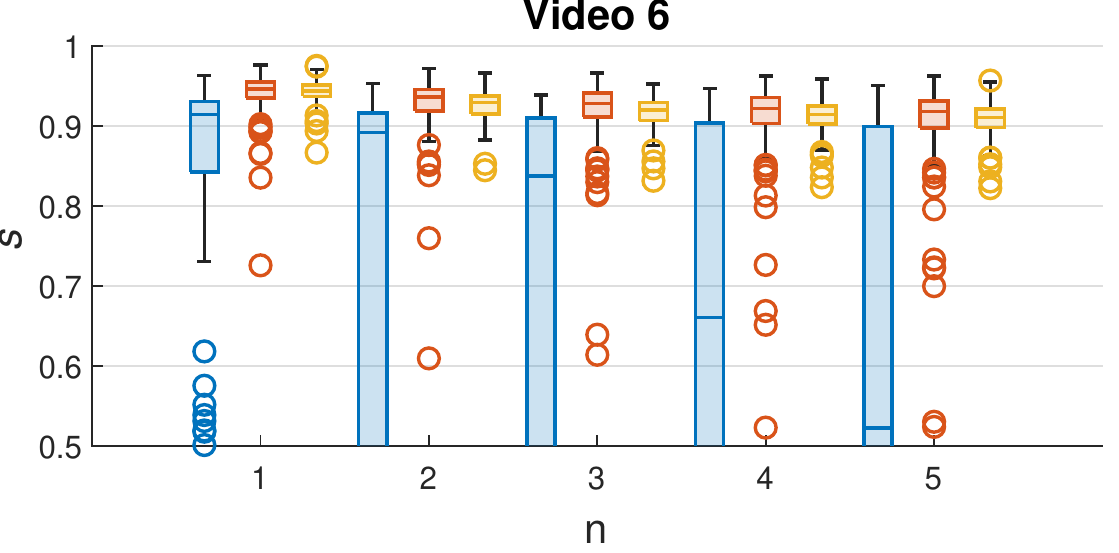} \\
\end{tabular}
}
\caption{Boxplots of $s$ over $n$ frames (with $n$ in range [1-5]) obtained with (blue) SIFT, (red)~\cite{bano2020deep} and (orange) the proposed framework.}
\label{fig:performance}
\end{figure*}

%To experimentally evaluate our hypothesis, \textcolor{red}{which hypothesis?} , 

%\textcolor{red}{quanto sono lunghi?} Table
%Representative sample frames from this dataset are shown in Fig.~\textcolor{red}{XX}.
%%
%

For qualitative evaluation, the registered frames are blended together using the Mertens-Kautz-Van Reeth exposure fusion algorithm~\citep{4392748} to tackle the non-uniform light exposure of the FoV along the fetoscopic video sequence.

%%%%%%%%%%%%%%%%%%%%

\subsection{Comparison with the literature and ablation study}

We compare our framework with SIFT, which is a standard feature extractor used for mosaicking by~\cite{daga2016real,reeff2006mosaicing}. 
We further compared our framework with with~\cite{bano2020deep}, which relies on deep learning for mosaicking and is the  best performing methods in the state of the art. 
For all our competitors, we replace any discarded homography with an identity matrix to preserve the frame numerical consistency across the methods. 
The ablation study characteristics are summarized in Table~\ref{tab:abl}.

As ablation study, we considered the following experiments:
\begin{itemize}
    \item Experiment 0 (E0): SuperPoint pre-trained on MS-COCO 2014 dataset, without any fine tuning on fetoscopy data.
    \item Experiment 1 (E1): Vanilla \textit{KPN}, as described in Sec.~\ref{kpc}. Here, both irrelevant keypoint rejection and inconsistent homography filtering are excluded. 
    \item Experiment 2 (E2): Improved \textit{KPN}, as described in Sec.~\ref{sec:improved_KPN}. Only inconsistent homography filtering is hence excluded.
\end{itemize}

For E2, we further investigates the performance obtained on an extended version of the dataset presented in Sec.~\ref{sec:dataset}. This extended version consists of the same 6 videos, but each video has an extended length (avg sequence length = 546 $\pm$ 237 frames). This allows us to evaluate the benefits of introducing homography filtering for longer video sequences.

 %, some artifacts could be produced in the frames blending. To improve the visualization of the mosaick, we apply the Mertens-Kautz-Van Reeth exposure fusion algorithm~\cite{4392748} in the rendering process.
%The Mertens-Kautz-Van Reeth exposure fusion is not applied for the computation of the metric.

\begin{figure*}[t!]
    \centering
    \makebox[\textwidth]{
\begin{tabular}{cc}
  \includegraphics[width=0.45\textwidth]{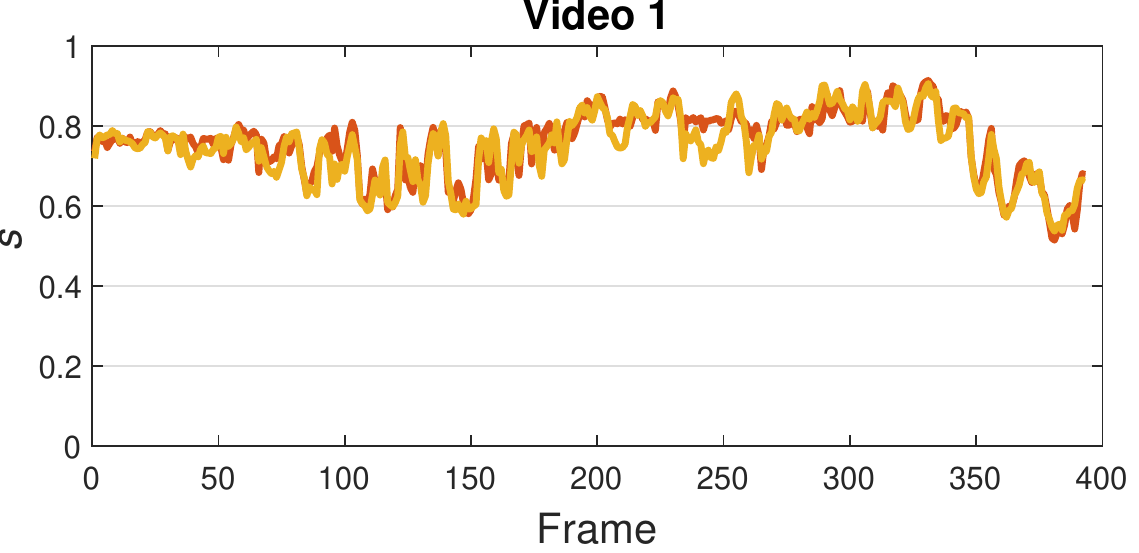} &   \includegraphics[width=0.45\textwidth]{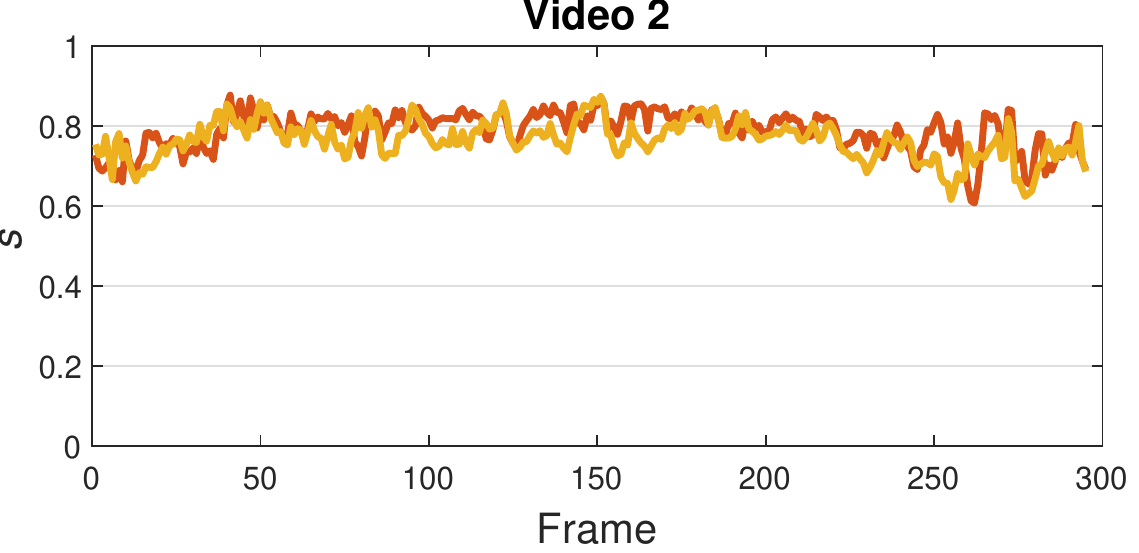} \\ \includegraphics[width=0.45\textwidth]{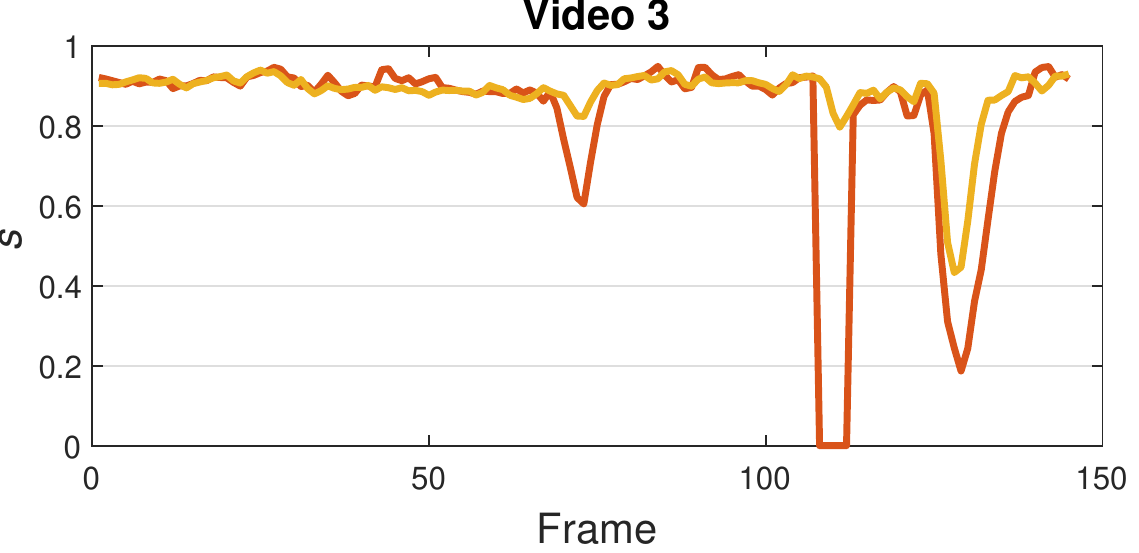} & \includegraphics[width=0.45\textwidth]{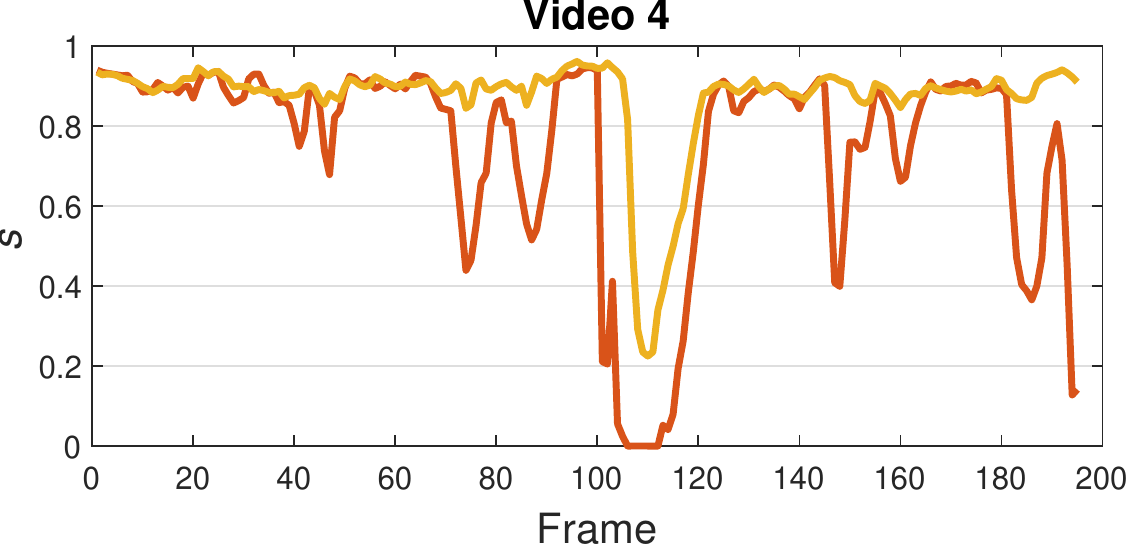} \\   \includegraphics[width=0.45\textwidth]{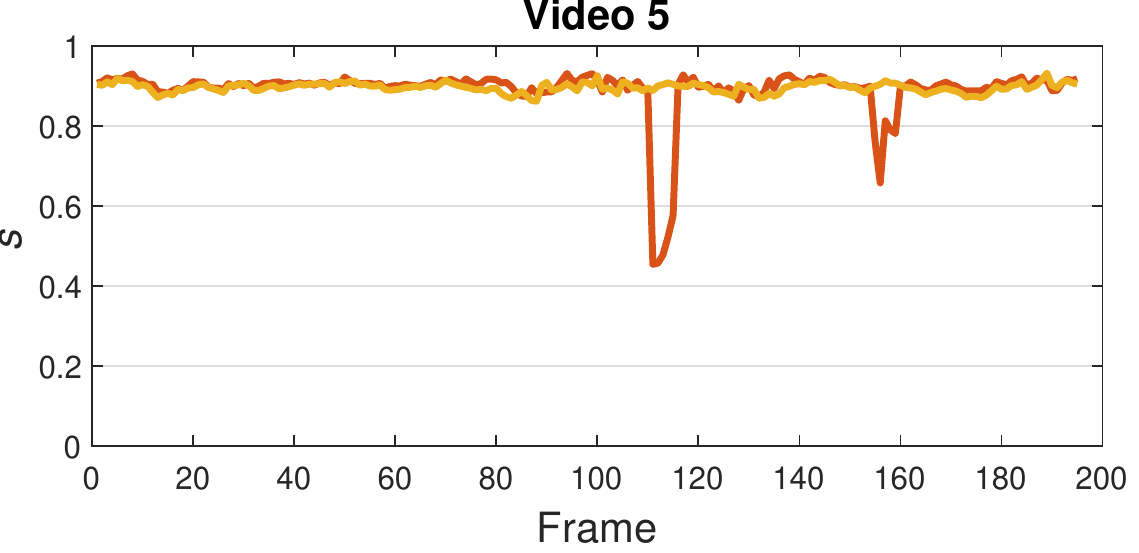} & \includegraphics[width=0.45\textwidth]{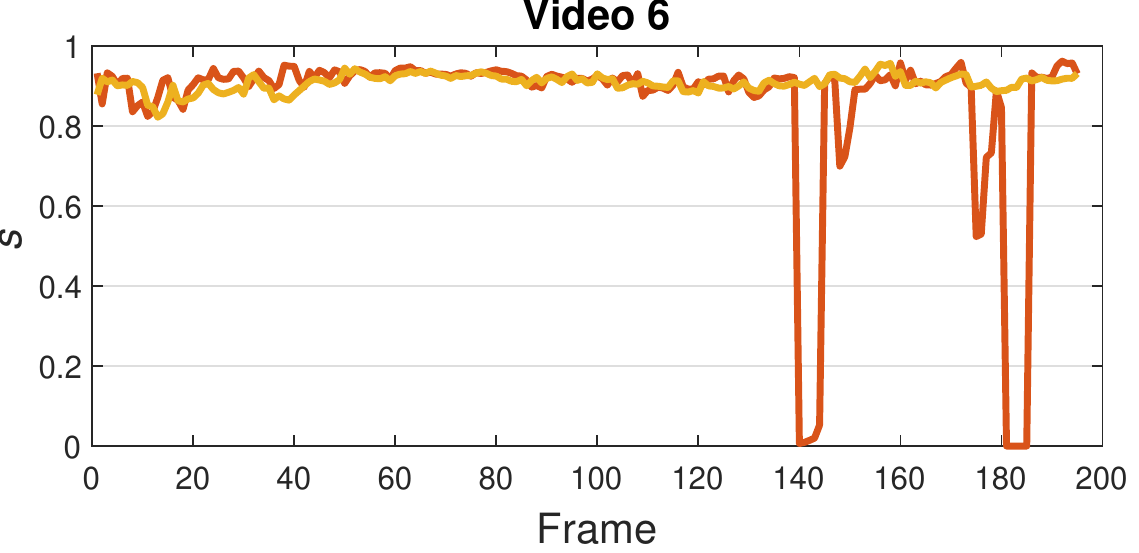} \\
\end{tabular}
}
    \caption{Plot of $s$ with $n$=5 computed for all video length. The curves refer to (red)~\cite{bano2020deep} and (orange) the proposed framework. }
    \label{fig:ssim}
\end{figure*}

\section{Results}\label{results}

The average $s$ values with $n$ equal to 5 obtained with SIFT, the work in~\cite{bano2020deep} and the proposed framework are reported in Table~\ref{tab:res}.
SIFT shows the lowest performance, as it fails in retrieving keypoints for mosaicking for several frames in all the 6 videos. This is in agreement with similar findings in the SDS/CAI field reported by~(\cite{liu2020extremely}). 

    For {Video~1} and {Video~2}, where vessels are clearly visible and lens distortion is small, we obtained $s$ with $n$=5 equal to $0.750 \pm 0.050$ and $0.766 \pm 0.048$, respectively. These results are comparable to that of~\cite{bano2020deep} ($0.757 \pm 0.081$ and $0.788 \pm 0.050$, respectively). Hence, the work in~\cite{bano2020deep} slightly outperforms the proposed framework for Video~1 and Video 2 by only $0.007$ and $0.022$, respectively. 
This was not true when considering the other videos, where the average $s$ was the highest for the proposed framework, which also granted the lowest standard deviation. The proposed framework overcomes~\cite{bano2020deep} by at least $0.007$ (video 5), with the highest difference for video 6 ($0.045$) and video 4 ($0.125$). %, achieving for  {Video~3} ($0.884 \pm 0.075$ and $0.839 \pm 0.208$), {Video~4} ($0.870 \pm 0.125$ and $0.745 \pm 0.257$) and {Video~6} ($0.909 \pm 0.021$ and $0.861 \pm 0.205$).

Figure~\ref{fig:performance} reports the value of $s$ at different $n$ obtained with SIFT, the work in~\cite{bano2020deep} and the proposed framework for the 6 tested videos. The proposed framework consistently outperformed the competitors for every $n$ for videos from 3 to 6. In the first two videos, the performance of the proposed framework were comparable to that of~\cite{bano2020deep}.

Figure~\ref{fig:ssim} shows the trend of $s$ with $n$ = 5 in time for the proposed method and~\cite{bano2020deep}. 
The trend of $s$ for~\cite{bano2020deep} in videos from 3 to 6 shows  drops in correspondence of wrong homography estimation. This happens to a lesser extent also for the proposed framework, but only in videos 3 and 4.

\begin{figure*}[t!]
    \centering
   \includegraphics[width=1.0\textwidth]{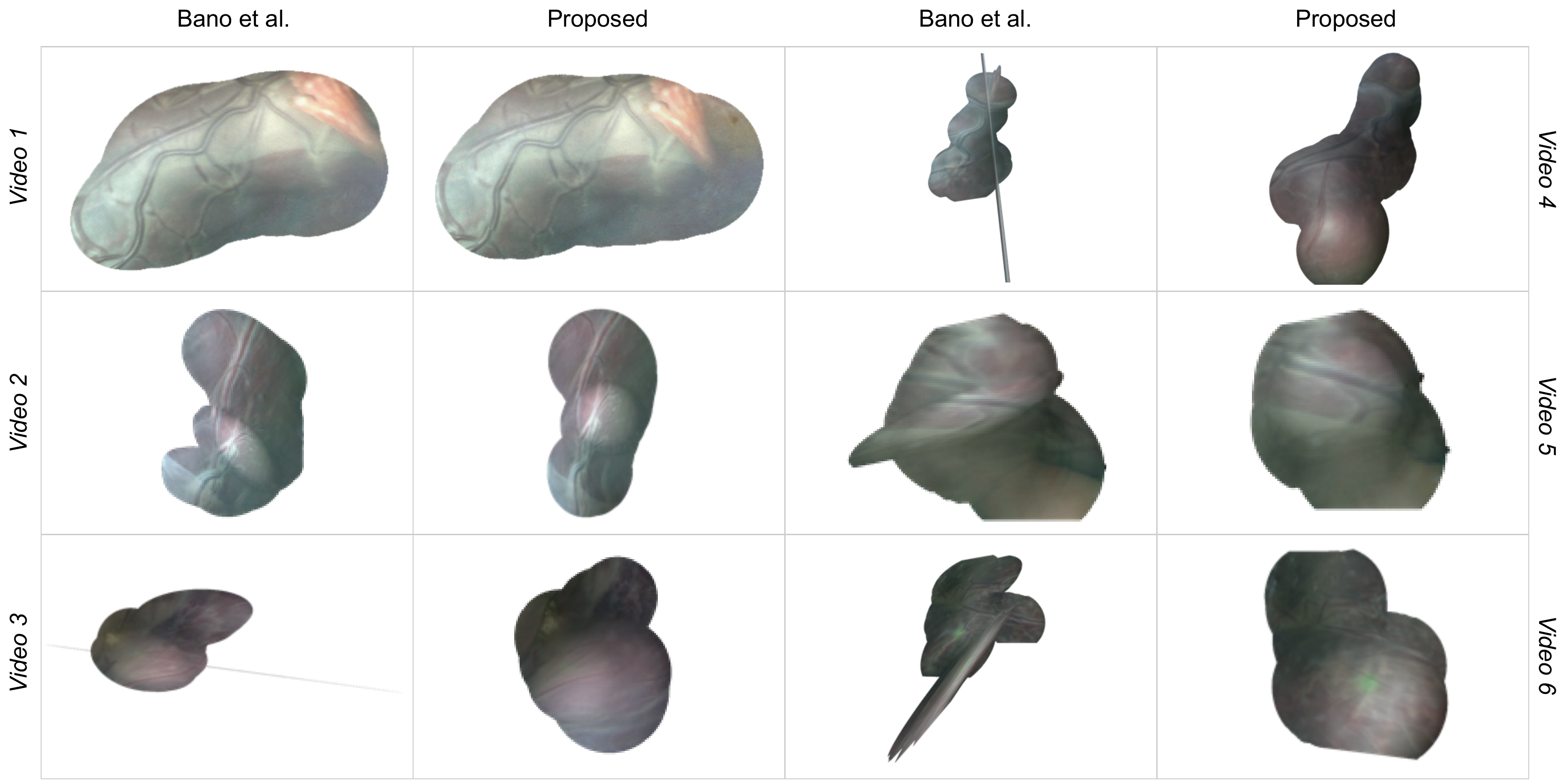}
    \caption{Mosaics obtained from the 6 TTTS videos using the method from~\cite{bano2020deep} and the proposed framework. Results refer to the dataset presented in Sec.~\ref{sec:dataset}.}
    \label{fig:qualitative}
\end{figure*}

The quantitative analysis presented in Fig.~\ref{fig:ssim} may also be appreciated from the qualitative examples shown in Fig.~\ref{fig:qualitative}, where the proposed framework achieves good-quality mosaicking for all the tested videos also when vessels are not visible.

The pretrained SuperPoint (E0) achieves $s$ over $n=5$ frame of $0.331$, showing lower performance than also SIFT.
E1, which aims at assessing the performance of vanilla \textit{KPN} alone, hence excluding both inconsistent keypoint rejection and homograghy filtering. In this experiment, we achieve an average $s$ of $0.788$, with a lost of $0.058$ over the proposed framework.
%and respectively $0.528 \pm 0.247$ {Video~1}, $0.202 \pm 0.264$ {Video~2}, $0.219 \pm 0.266$ {Video~3}, $0.322 \pm 0.362$ {Video~4}, $0.341 \pm 0.284$ {Video~5}, and $0.209 \pm 0.336$ {Video~6}.

%
With our ablation study E2, which aims at evaluating the benefit of introducing inconsistent keypoint rejection after the \textit{KPN}, we achieve an average $s$ with $n=5$ of $0.848$. 
Despite the relatively small difference (0.064) in the performance achieved by our framework over E2, inconsistent homography filtering allows us to lower the drift in the mosaic and mitigate tracking loss in challenging videos, where images are strongly under-exposed or whether noisy keypoints are computed (e.g., due to particles). %This can also be observed from the visual results in the supplementary material.
%All these experiments were performed on the dataset by [4] (6 videos from 6 patients, 1450 frames, avg sequence length = 237 ± 92 frames) for a fair comparison with state of the art.
%As for E2, we does not find any benefits when processing the dataset presented in \cite{bano2020deep}. 
However, when processing the extended version of this dataset with longer sequences, our results improve by $3\%$ when adding homography filtering, as shown in Table~\ref{tab:res_ext}.

\begin{table*}[t!]
\centering
\caption{Quantitative results for the extended dataset with longer fetoscopy sequences.}  \label{tab:res_ext}
\begin{adjustbox}{max width = \textwidth}
\begin{tabular}{cccc}
                           & Video 1      & Video 2      & Video 3      \\ \hline
\multicolumn{1}{l}{E2} & $0.735 \pm 0.154$ & $0.710 \pm 0.014$ & $0.811 \pm 0.210$ \\

\multicolumn{1}{l}{Proposed} &  $0.751 \pm 0.098$ & $0.771 \pm 0.072$ & $0.886 \pm 0.091$
\\ \\
                           & Video 4      & Video 5      & Video 6      \\ \hline
\multicolumn{1}{l}{E2} &  $0.810 \pm 0.140$ & $0.802 \pm 0.320$ & $0.791 \pm 0.164$\\

\multicolumn{1}{l}{Proposed} & $0.872 \pm 0.132$ & $0.896 \pm 0.022$ & $0.901 \pm 0.051$
 \\
\end{tabular}
\end{adjustbox}
\end{table*}

\begin{figure}[t!]
\centering
    %\adjust[\textwidth]{
\begin{tabular}{ccc}

  \includegraphics[width=.25\textwidth]{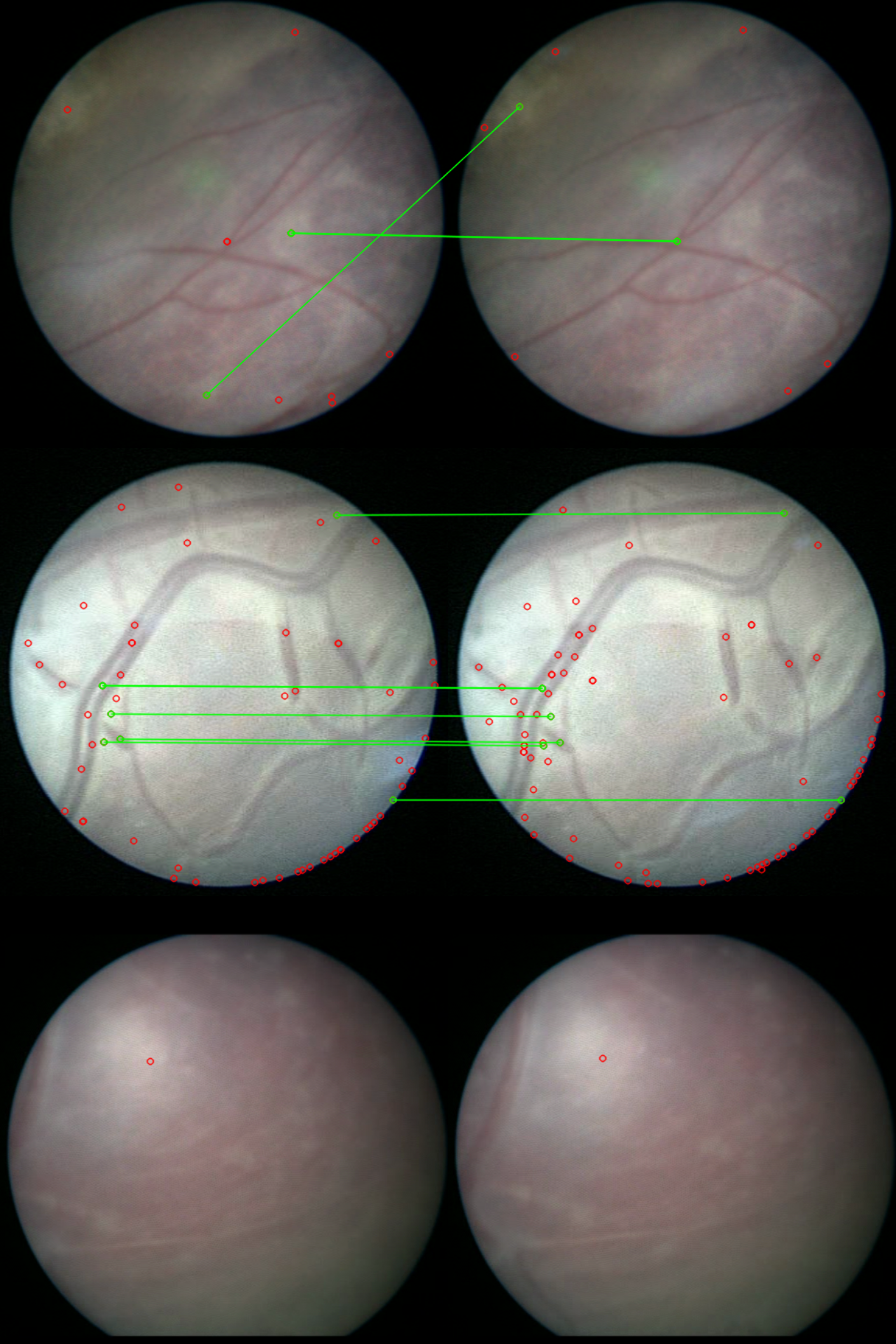} & \includegraphics[width=.25\textwidth]{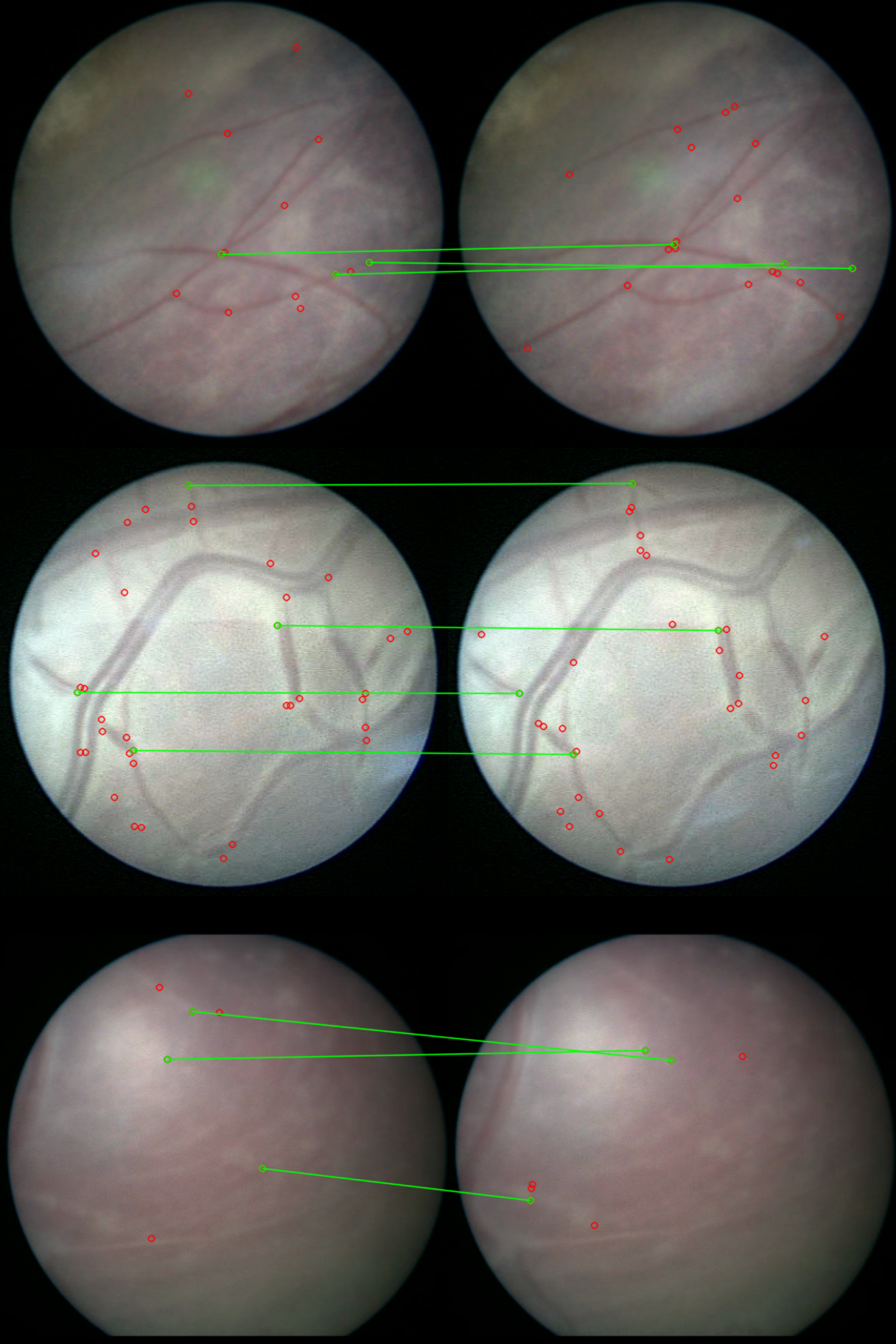} & \includegraphics[width=.25\textwidth]{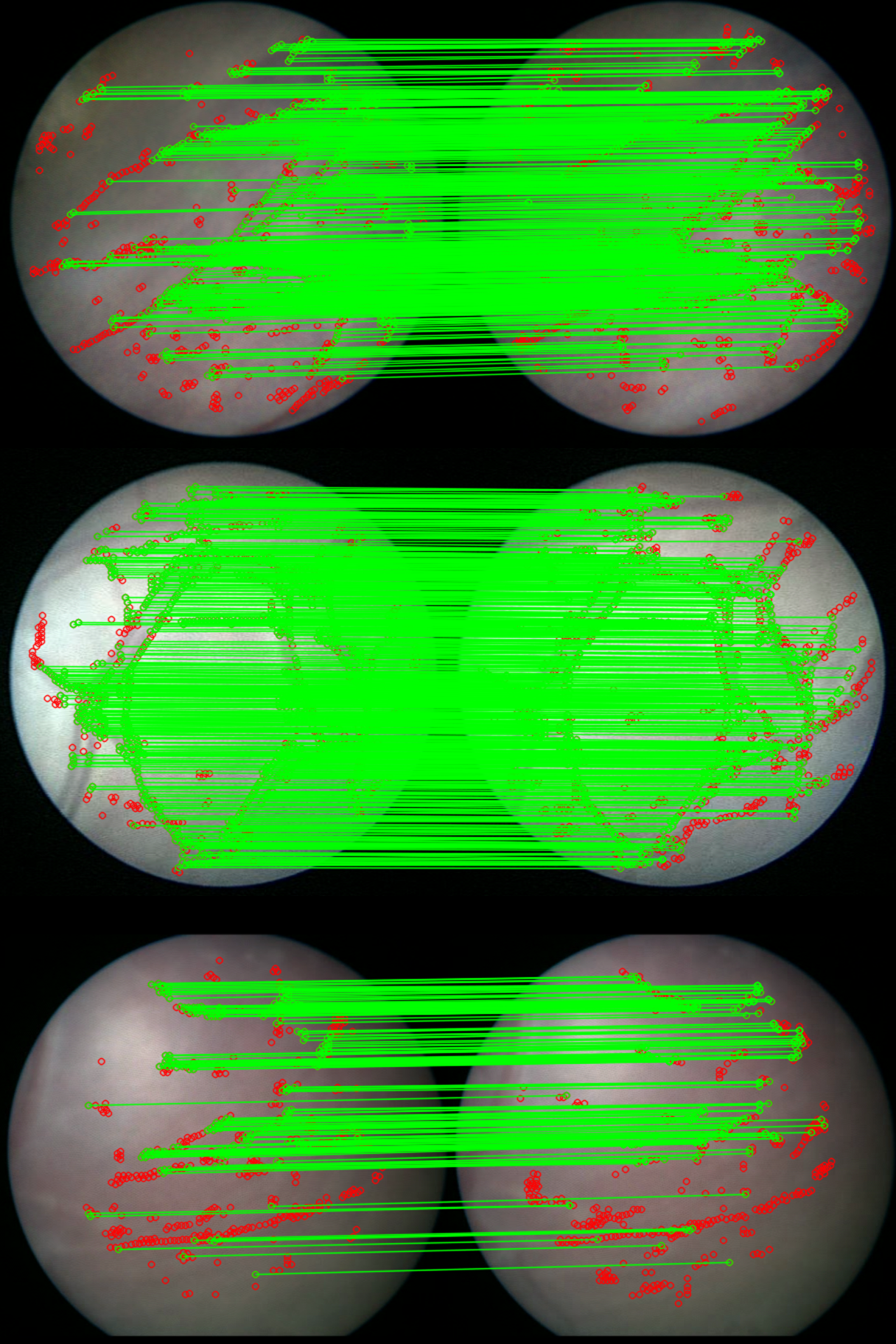} \\
\end{tabular}

%}
\caption{Visual comparison of keypoint computation from (left) SIFT, (middle) E0 and (right) proposed framework. }
\label{fig:kp_visual_comparison}
\end{figure}

% . The estimation of keypoints on these structures can bias the estimation of camera movement.
%Furthermore, the computed keypoints lies on placenta surface as well  but also on placenta surface.
%These findings could explain why the proposed method achieved better performance in the other sequences and is more robust to sudden motion or distortions. 

%While promising results have been achieved for mosaicking from short video sequences, long-term mapping still remairins an open challenge. This may be attributed to the high intra- and inter-case variability in each procedure, dynamically challenging non-planar views, poor visibility, texture paucity, low resolution and occlusion due to the presence of fetus and ablation tool in the field-of-view.  

\section{Discussion and conclusions}\label{disc}
In this work, we proposed a mosaicking framework to perform FoV expansion in fetoscopy videos using learning-based keypoints. Going beyond the current state of the art, our framework does not rely on any prior vessel segmentation for keypoint prediction, which makes it robust when registering frames where vessels are not clearly visible or when vessel segmentation is not accurate. We instead use surgical-tool and fetus segmentation to filter out irrelevant keypoint and propose a simple yet effective strategy to filter out unrealistic homographies. 
%When evaluated on 6 different in-vivo fetoscopic videos, our framework showed improved performance over the current state of art, paving the way for the development of TTTS navigation systems to support surgeons in TTTS laser-ablation procedures.

To test our first research hypothesis (H1), we applied our proposed framework on challenging clinical videos from 6 different TTTS surgeries. We also compared the proposed framework with state-of-the-art approaches for fetoscopy mosaicking (Table~\ref{table:dataset}, showing that our method performs well when others fail. From our experiments, as shown in Fig.~\ref{fig:kp_visual_comparison}, SIFT was not always able to compute a sufficient number of keypoints to compute homography. When keypoints were computed in a sufficient number, they did not allow us to model camera motion. This can be explained considering that SIFT is not robust in case of images with low contrast and texture, as is the case of fetoscopic frames.

As for the comparison with~\cite{bano2020deep}, the absence of placenta vessels in a number of consecutive frames ({Video~3}, 4 and 5) compromised the registration process, while this was not the case for our framework.
Moreover, from {Video~3} to {Video~6} the placenta surface is not perfectly planar in all frames, the lens distortions is more evident and camera moves along different planes to scan the entire placenta surface. Nonetheless, the proposed framework did not fail in providing good quality mosaicking.
This can be explained considering our self-supervised training with homographic augmentation, which allowed us to detect robust keypoints to estimate homography despite changes in perspective.
Our framework hence allows FoV expansion also in videos that suffer from change of perspective, as in case of anterior placenta ({Video~3} to {Video~5}).

With our ablation study (E0), we further highlighted the benefit of the proposed framework over the pre-trained SuperPoint without any fine tuning on fetoscopy data. The pre-trained SuperPoint has been optimized to compute robust keypoints from natural images, and thus face challenges in dealing with the complexity of fetoscopic images. This explain why, its out-of-the shelf performance is even lower than SIFT.
%\hl{However, the need of fine tuning a learning-based method is related to its capatability to adapt keypoints computation from specific domain features.}

Our second research hypothesis (H2) was focused on assessing the benefits of including irrelevant keypoint rejection and inconsistent homography filtering. 
When analyzing $s$ over the entire sequences (Fig.~\ref{fig:ssim}), our framework showed a lower number of drops in $s$ than~\cite{bano2020deep}.
However, small drops were present in {Video~3} and {Video~4}.
This may be due to underexposed frames where keypoint estimation is particularly challenging. However, as the amount of underexposed frames was reasonably small, the inconsistent homography filter was able to tackle the challenge.

The benefit of adding inconsistent homography filtering was specifically useful in long range videos from the extended dataset.%, as shown in the supplementary materials. %\hl{Fig. }. 
We explain this improvement considering that the extended dataset includes further challenges (i.e., field of view occlusions, faster fetoscope movements and extreme change in illumination). 

The proposed approach allows for the computation of robust keypoints from intra-operative images, which, contrary to methods proposed in the literature, allows for fetoscopic mosaics with less drift. 
As an additional advantage, obtaining keypoints and descriptors enables the integration with localization and mapping frameworks (e.g., SLAM) widely employed for robotic and automotive, paving the way towards the realisation of a complete mosaicking and navigation system for fetal surgery.

Possible limitations of the proposed framework may be encountered during sudden changes in illumination or with highly over- or under-exposed images. 
In these circumstances, it may not be possible to detect a sufficient number of keypoints to calculate homography robustly. In such a case, the inconsistent homography filter may mitigate the failure of mosaicking only if the changes happen within a few frames.
Another possible limitation of this framework is the absence of maternal breath handling. Although this does not compromise the usability of the generated mosaic, it may introduce some minor distortions. 

A limitation of the experimental protocol may be seen in the dataset size, but this is currently the largest available dataset for in-vivo fetoscopy mosaicking. The proposed framework may also be potentially translated to other surgical fields, including neuro microsurgery for the treatment of gangliogliomas. 

Our experimental results suggest that the proposed framework may be effective in supporting surgeons during surgical procedures for treating TTTS by providing a larger FoV. 
This may have a positive impact, by reducing surgeons’ mental workload and, as a consequence, potentially reducing patients’ risks and lowering surgery duration.

\section*{Ethical standards}

The proposed study is a retrospective study. Data used for the
analysis were acquired during actual surgery procedures and then
were anonymized to allow researchers to conduct the study. All the
patients gave their consent on data processing for research purpose. The study fully respects and promotes the values of freedom,
autonomy, integrity and dignity of the person, social solidarity and
justice, including fairness of access. The study was carried out in
compliance with the principles laid down in the Declaration of
Helsinki, in accordance with the Guidelines for Good Clinical Practice.

\section*{Declaration of Competing Interest}
No benefits in any form have been or will be received from a
commercial party related directly or indirectly to the subjects of
this manuscript.

%%Harvard
\bibliographystyle{model2-names.bst}\biboptions{authoryear}
\bibliography{refs}

%\section*{Supplementary Material}

%Supplementary material that may be helpful in the review process should be prepared and provided as a separate electronic file. That file can then be transformed into PDF format and submitted along with the manuscript and graphic files to the appropriate editorial office.

\end{document}